# Optical characteristics and capabilities of the successive versions of Meudon spectroheliograph (1908-2023)

J.-M. Malherbe, Observatoire de Paris, PSL Research University, CNRS, LESIA, Meudon, France

Email : Jean-Marie.Malherbe@obspm.fr

ORCID id : https://orcid.org/0000-0002-4180-3729

Date : 19 March 2023

## ABSTRACT

The spectroheliograph is a spectroscopic instrument designed to produce monochromatic images of the photosphere (the visible layer) and the chromosphere of the Sun. It was invented at the same time (1892), but independently, by Hale in the USA and Deslandres in France and was dedicated to long-term surveys of the solar cycles. For that purpose, systematic observations of the CaII K and Hα lines started in Meudon observatory in 1908 and continue today, so that a huge collection of more than 100000 spectroheliograms, spanning 115 years of solar activity, was recorded. We present in this paper the optical characteristics and the capabilities of the successive versions of the instrument, from 1908 to now.

## INTRODUCTION

Henri Deslandres (1853-1948) was hired in 1889 by Admiral Ernest Mouchez (the director of Paris observatory) and was in charge of organizing a spectroscopic laboratory, in the context of the development of physical astronomy initiated by Jules Janssen (1824-1907) at Meudon. Deslandres first built a classical spectrograph and studied the profiles of the CaII K line at 3934 Å wavelength; he succeeded to resolve the fine structure of the core in 1892. This was the starting point of the spectroheliographs in France (Malherbe, 2023a). With a classical spectrograph, monochromatic images can be produced by an output slit located in the spectrum (to select the light of a spectral line) when the input slit scans the solar surface. This is possible by moving the solar image upon the first slit with the translation of the imaging objective. The photographic spectroheliograph was invented on this basis by George Hale in Kenwood (1892) and by Henri Deslandres in Paris (1893), simultaneously but independently. The principle is explained by d'Azambuja (1920a, 1920b).

Many spectroheliographs were constructed in the world to study the photosphere (thin layer of 500 km with temperature decreasing from 6000 K to 4500 K and exhititing magnetic structures, such as sunspots and bright faculae) and the chromosphere above (layer of 2000 km thickness, with temperature increasing from 4500 K to 8000 K, revealing structures such as dark filaments and bright plages). The chromosphere requires spectroscopic means to be seen. Long series of continuous observations in the CaII K line were collected by the spectroheliographs of Kodaikanal in India, Mount Wilson in the USA, Mitaka in Japan, Sacramento Peak in the USA, Coimbra in Portugal, Meudon in France or Arcetri in Italy. They produced extended archives, some of them covering up to 10 solar cycles, which are convenient to investigate long-term solar activity.

Spectroheliographs historically produced monochromatic images along solar cycles at the epoch of photographic plates. Most of them observed the CaII K and Hα Fraunhofer lines of the chromosphere (Malherbe, 2023b). They were progressively abandoned in the second half of the twentieth century for telescopes using narrow bandpass filters (such as Fabry-Pérot or Lyot filters), which are more compact and able to observe, at much higher cadence, dynamic events (such as flares). However, with modern and fast electronic detectors, there is a regain of interest for imaging spectroscopy, because digital spectroheliographs can now deliver not only monochromatic images, but also full line profiles for each pixel of the Sun.

Section 1 presents the optical capabilities of the initial and large quadruple spectroheliograph designed in 1908 by Deslandres and Lucien d'Azambuja (1884-1970), while section 2 describes the three successive versions of the modern era (1989-2023). Section 3 is dedicated to the SOLar Explorer (SOLEX) instrument, a collaborative project between amateurs and professional astronomers for observations with a mini (but high performance) spectroheliograph, in order to increase the number of annual observations.

# 1 – THE LARGE QUADRUPLE SPECTROHELIOGRAPH OF DESLANDRES AND D'AZAMBUJA

Deslandres moved from Paris to Meudon in 1898 and installed there the first versions of his spectroheliographs, but observations were not yet systematic and many experiments and tests were performed to improve the quality of monochromatic images. A new building (figure 1) was erected in 1906 for the installation of a large quadruple spectroheliograph, wich was built in collabotation with d'Azambuja, who was hired in 1899, and was also in charge of organizing the future service of systematic observations. Hence, daily observations of CaII K started in 1908 and were followed soon by Hα in 1909, and continue today. They were just interrupted during four years by WW1.

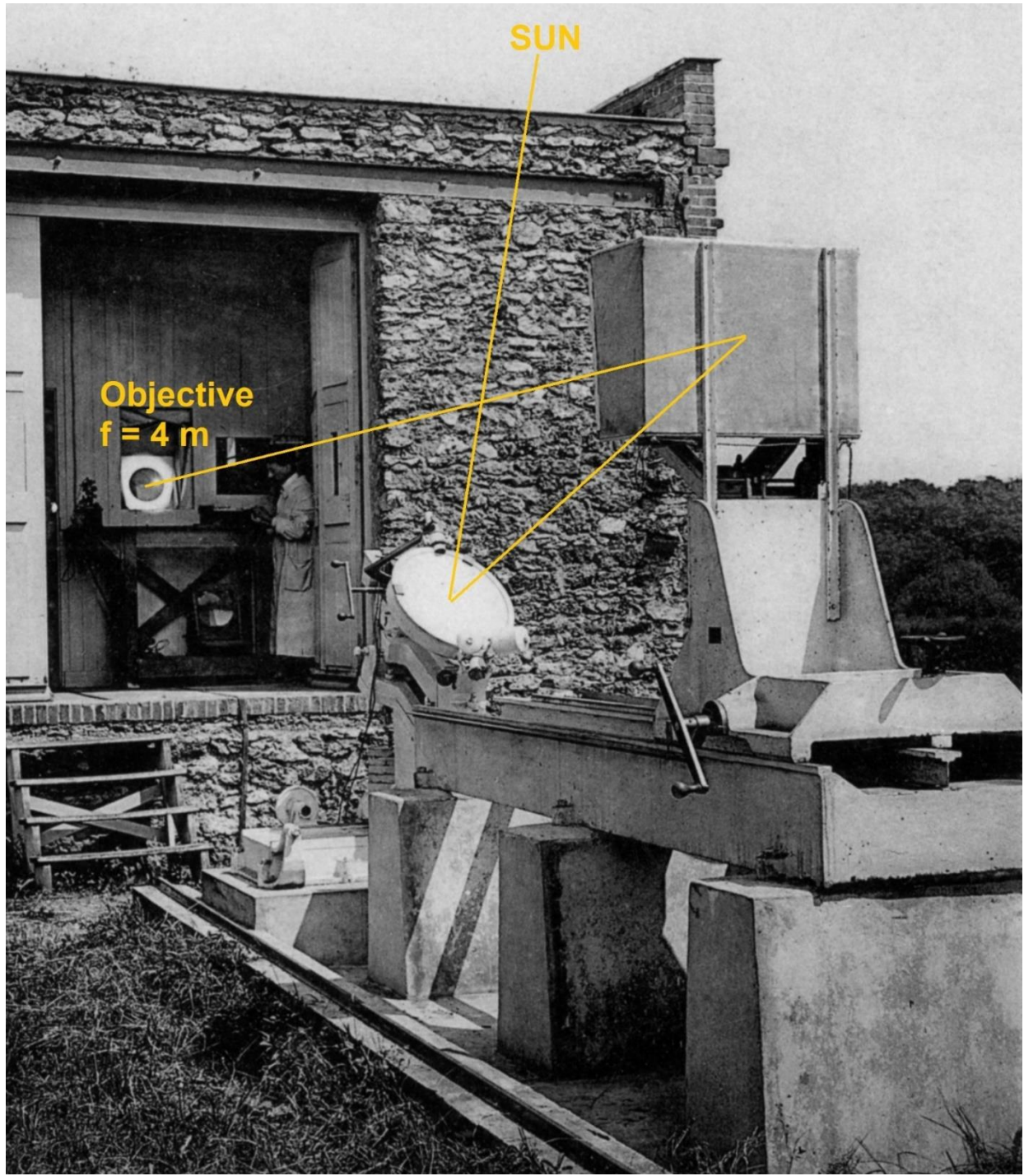

**Figure 1**. *The coelostat of the spectroheliograph catches the solar light and redirects it in the horizontal and Nort/South direction of the laboratory. It uses two 40 cm flat mirrors. The first one follows the Sun (rotation speed of one turn during 48 hours); the second one reflects the beam towards the 4-metres focal length entrance objective (250 mm diameter). Courtesy Paris Observatory.*

The large spectroheliograph was horizontal and stayed in a fixed position. For that reason, it was fed by a coelostat (two flat mirror system, North of the building) as shown by figure 1. The instrument was composed of four spectroheliographs (figure 2) that could not operate simultaneously, but alternatively:

1 – spectroheliograph n° I: a 3-metres instrument dedicated to systematic observations in CaII K line, with three flint prisms of 60°. The beam in the CaII K chamber is indicated in violet. The core of the line (K3) or the violet wing (K1v, -1.3 Å from the line core) were selected successively.

2 – spectroheliograph n° II: a 3-metres instrument dedicated to systematic observations in Hα line, with a plane grating (568 grooves/mm). The beam in the Hα chamber is indicated in red.

3 – spectroheliograph n° III: a multi-purpose 3-metres instrument, using either the three prisms or the grating.

4 – spectroheliograph n° IV: a research 7-metres high dispersive instrument using either the three prisms or the grating, with an associated afocal reductor (n° IVbis). The beam in the chamber (at right) and the reductor (at left) is indicated in blue. The photographic plate can be put either in $F_4$ (no reduction) or in f (with reduction). The different combinations are summarized by table 1. They all use a common entrance objective (4.0 m focal length), collimator (1.30 m focal length) and entrance slit F (0.030 mm = 1.55" width).

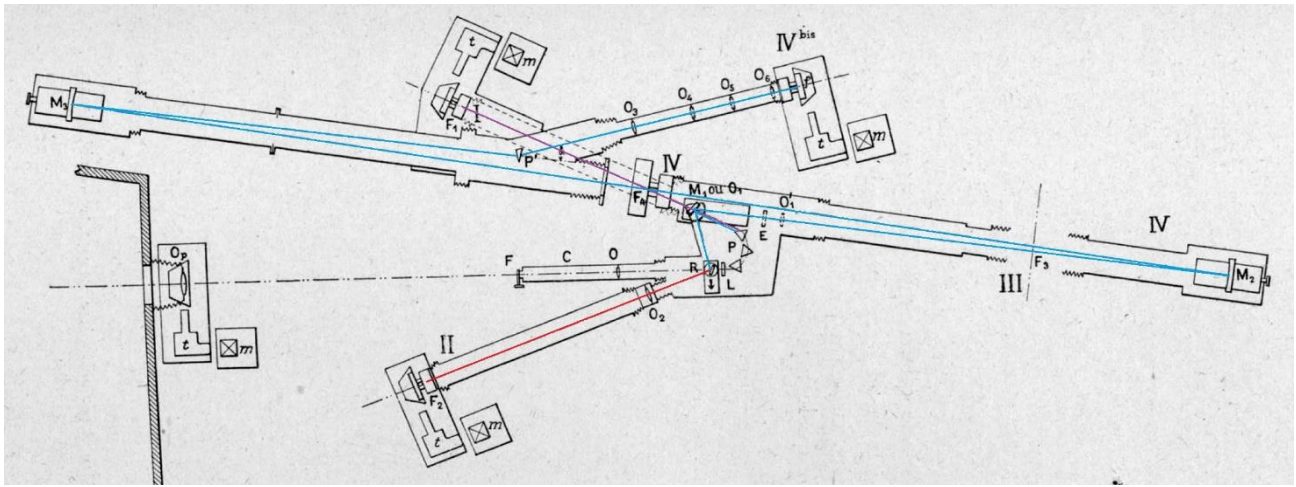

**Figure 2**. *The optical combinations of the quadruple spectroheliograph. The entrance objective (Op, 4.0 m focal length), slit (F) and collimator (O, 1.3 m focal length) are common to the four chambers n° I (violet beam), n° II (red beam), n° III and n° IV (blue beam). (t, m) are couples of velocity transformers and motors driving the 13 x 18 cm² glass plates. R and P are the disperser devices (grating or 3 prisms). L is a plane-cylindrical lens (astigmatism corrector of prisms). $F_1$, $F_2$, $F_3$, $F_4$, f are slits located in the spectrum. $O_1$, $O'_1$, $O_2$ are chamber lenses (3.0 m focal length) while $M_2$ is a concave mirror (7.0 m focal length) respectively for combinations n° I, n° III, n° II and n° IV. $M_3$ and either $O_3$, or $O_4$, or $O_5$, or $O_6$ constitute an afocal reducer (concave mirror of 7.0 m focal length and lenses of various power) in order to decrease the image size of the combination n° IV. They form together the combination n° IVbis. $M_1$ is a removable flat mirror used for spectros n° III and n° IV, while P' is a 30° prism to separate grating orders and parasitic light of spectro n° IV. After d'Azambuja (1930).*

**Table 1**. *The optical combinations of the quadruple spectroheliograph of figure 2. After d'Azambuja (1930).*

| Chamber | Optical Path (figure 2) | Length (m) | Image diameter (mm) | Spectral line |
|---|---|---|---|---|
| **For systematic observations** | | | | |
| n° I | P+$O_1$+$F_1$ | 3.0 | 86 | CaII K |
| n° II | R+$O_2$+$F_2$ | 3.0 | 86 | Hα |
| **For research observations** | | | | |
| n° III | (P or R)+$M_1$+$O'_1$+$F_3$ | 3.0 | 86 | multi-purpose |
| n° IV | (P or R)+$M_1$+$M_2$+$F_4$ | 7.0 | 205 | High resolution |
| n° IVbis (n° IV with | (P or R)+$M_1$+$M_2$+$F_4$+$M_3$+P' | 2 x 7.0 | | |
| image reduction γ) | with $O_1$    (γ = 0.29) | | 58.5 | High spectral |
| | or $O_2$     (γ = 0.19) | | 39.5 | resolution |
| | or $O_3$     (γ = 0.09) | | 19.5 | with image |
| | or $O_4$     (γ = 0.04) | | 9.0 | reduction |

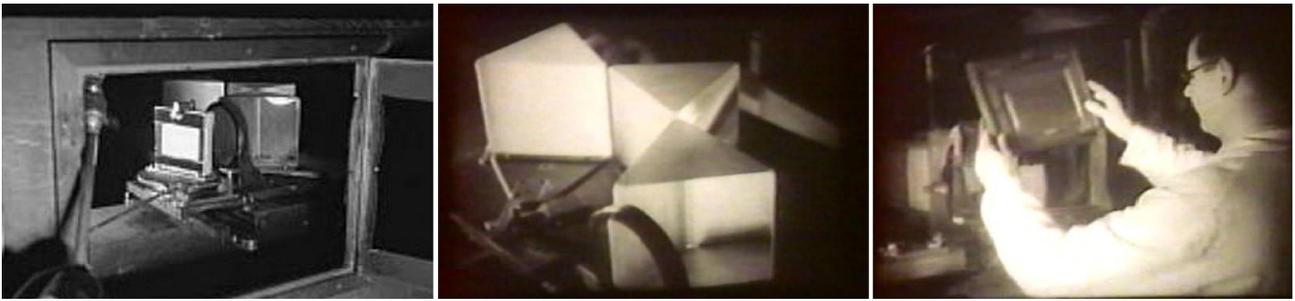

**Figure 3.** *The plane grating (568 grooves/mm), flint prisms (60°) and photographic plate support. Courtesy Paris Observatory.*

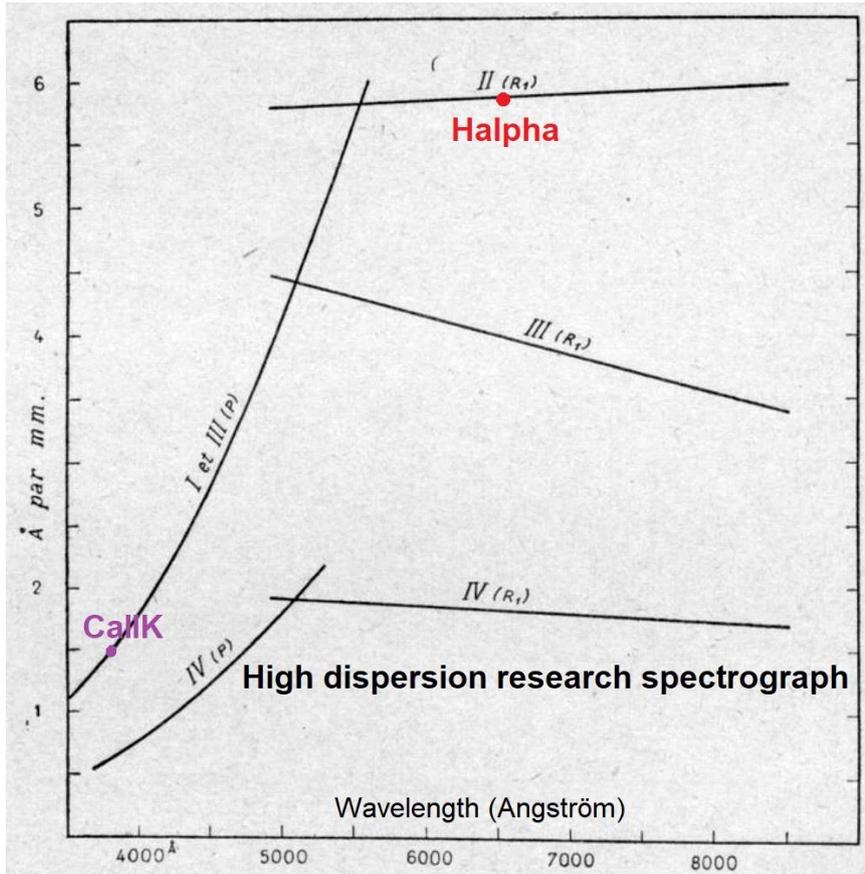

**Figure 4.** *The dispersion in Å/mm of the quadruple spectroheliograph.*

*Three 3-metres chambers:*
*n° I: 3 prisms, for CaII K systematic observations.*
*n° II: plane grating, for Hα systematic observations.*
*n° III: 3 prisms or plane grating, multi-purpose, medium dispersion, for research.*

*The 7-metres chamber:*
*n° IV: 3 prisms or plane grating, high dispersion for research observations.*

*After d'Azambuja (1930).*

**Table 2.** *Characteristics of the four spectrographs in 1909. After d'Azambuja (1930).*

| Spectrograph number | Spectrograph type | Focal length (m) | Spectral range (Å) | at λ (Å) | Dispersion (Å/mm) with prisms or grating | | Image diameter (mm) |
|---|---|---|---|---|---|---|---|
| n° I | 3 prisms | 3.0 | 3900-4100 | 3934 | 1.7 | | 86.5 |
| n° II | grating | 3.0 | 3600-9000 | 3934 | | 5.7 | 86.5 |
| | | | | 5184 | | 5.8 | 87.5 |
| | | | | 6563 | | 5.9 | 89.5 |
| n° III | prisms/grating | 3.0 | 3600-9000 | 3934 | 1.7 | 4.8 | 86.5 |
| | | | | 5184 | 4.6 | 4.4 | 87.5 |
| | | | | 6563 | 9.4 | 4.0 | 89.5 |
| n° IV | prisms/grating | 7.0 | 3600-9000 | 3934 | 0.7 | 2.0 | 206.0 |
| | | | | 5184 | 2.0 | 1.9 | 205.5 |
| | | | | 6563 | 4.1 | 1.8 | 205.0 |

The prisms (13 x 15 cm², angle A = 61°, refractive index n = 1.65) and the plane grating (8 x 5.5 cm², 568 grooves/mm) are displayed in figure 3.

The set of 3 prisms, composed of flint glass, worked at the minimum deviation (d=51°/prism giving a full deviation of D = 3 d = 153°) and was used in the blue/violet part of the spectrum. In the CaII K line, for the flint, we have $dn/d\lambda$ = -0.355 $\mu^{-1}$, which provides the quantity $dD/d\lambda$ = 6 $(dn/d\lambda)$ sin(A/2) / cos((A+d)/2) = -1.88 $\mu^{-1}$ for the set of three prisms. The output slit in the spectrum had the width w = 0.075 mm, which corresponds to a small deviation dD = w/f (f = 3.0 m is the focal length of chamber n° I) = 2.5 $10^{-5}$ rd, from which we obtain the waveband of the slit $d\lambda$ = 0.13 Å. This value is in full agreement with figure 4 and table 2, which give the dispersion of 1.7 Å/mm in the CaII K line, providing the waveband $d\lambda$ = 1.7 w = 1.7 x 0.075 = 0.13 Å.

The plane grating was used in the first order and the angle between the incident and diffracted beam was 20°. The grating formula predicts a dispersion of 5.5 Å/mm for the 3.0 m chamber n° II used for Hα, in agreement with figure 4 and table 2, giving 5.8 Å/mm. The bandpass of the slit (w = 0.075 mm) is 0.43 Å.

In conclusion, the spectrographs n° I and n° II (figure 5) provided monochromatic images in CaII K and Hα with respectively 0.13 Å and 0.43 Å bandpass.

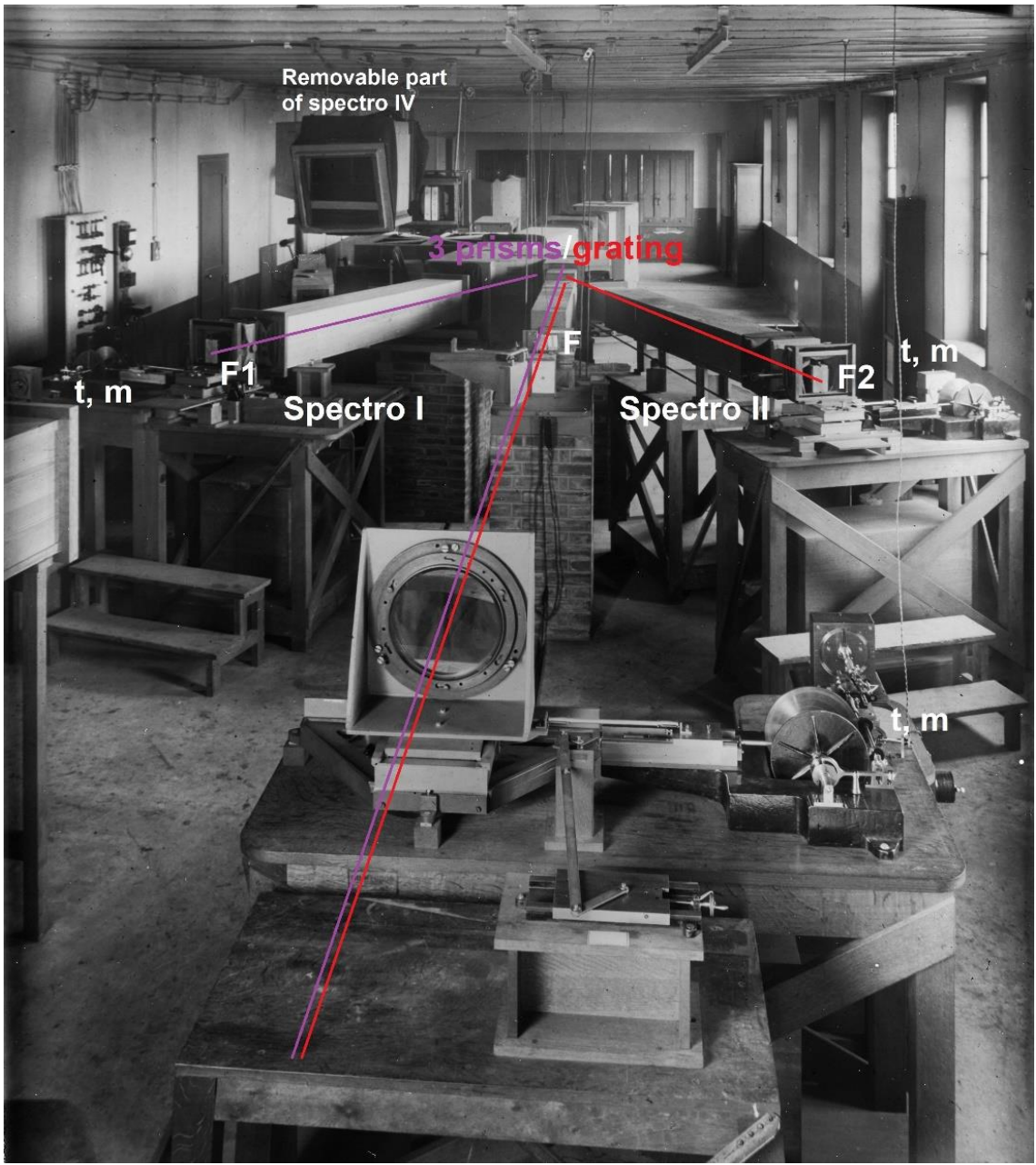

**Figure 5.** *Spectroheliographs n° I (3 prisms) and n° II (plane grating) providing alternatively spectroheliograms in the CaII K and Hα lines. $F_1$ and $F_2$ are the 12 cm output slits in the spectrum, in the focal plane of the chambers. (t, m) are couples of velocity transformers and motors which synchronize the translations of the 4.0 m focal length imaging objective and the photographic plates of the chambers. Courtesy Paris Observatory.*

Figure 6 shows two of the first spectroheliograms obtained in CaII K with chamber n° I (1908) and in Hα with chamber n° II (1909). Systematic observations started in 1908 and continue today, with an interruption during WW1.

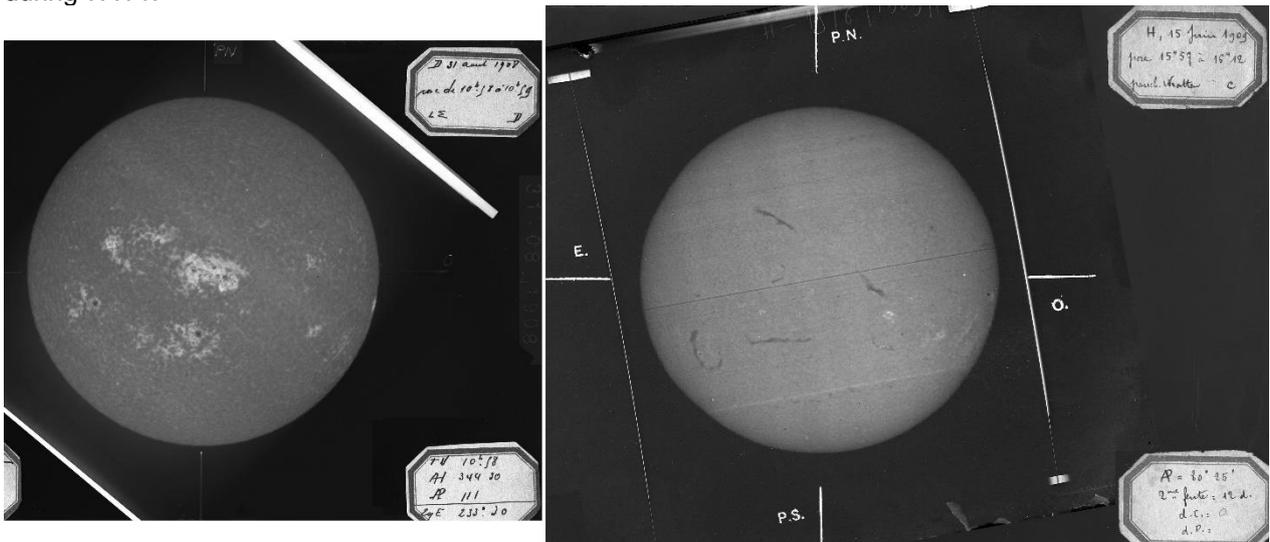

**Figure 6.** *Two examples of the first spectroheliograms obtained in CaII K (left, 1908) and Hα (right, 1909) in the frame of the systematic observations program. Courtesy Paris Observatory.*

The 7.0 metres chamber n° IV, dedicated to high dispersion spectroscopy and scientific research, was much more powerful and complicated than the others. For wavelengths smaller than 5000 Å, the set of 3 prisms was used, because in this range, the refractive index varies much with the wavelength, providing a dispersion in the range 0.5 - 1.5 Å/mm. But for the wavelengths above 5000 Å, the plane grating appeared better, and the dispersion was in the range 1.5 - 2.0 Å/mm.

The spectrograph used a 7.0 m focal length concave mirror (40 cm diameter) for the chamber and the spectrum formed at slit $F_4$ (figure 2). A 0.075 mm slit width corresponds there to a typical bandwidth of 0.10 Å, providing an excellent selectivity along the line profiles, so that chamber n° IV was convenient both for thin photospheric lines and broad chromospheric lines of the Fraunhofer spectrum. However, the magnification was so high (7.0/1.3 = 5.4) that the size of the solar image was 37.2 x 5.4 = 200 mm at $F_4$, too large to record the full Sun with usual glass plates of 13 x 18 cm². For that reason, $F_4$ was convenient only for regions.

In order to reduce the image size of the solar disk, an afocal reductor was added. It was composed of an afocal system with a second 7.0 m focal length concave mirror and a lens (chosen among four possibilities) to produce an image between 9 mm and 58 mm diameter (table 1). A 30° prism (P') was included in the optical path to reduce parasitic light and residual superimposed grating orders. In the focal plane, the output slit f had a usual width of 1 mm (because the waveband was always selected in $F_4$).

The curvature of the spectral lines R is roughly equal to (f²/λ) D, where f is the focal length of the chamber, λ the wavelength and D the dispersion (dλ/dx in Å/mm). Curved slits at $F_4$ were used. For this spectrograph, R was found to be (figure 7):
- with prisms:   4000 Å < λ < 5700 Å  →  2.10 m < R < 2.50 m (increasing with wavelength)
- with the grating: 5700 Å < λ < 8500 Å  →  2.50 m > R > 2.35 m (decreasing with wavelength)

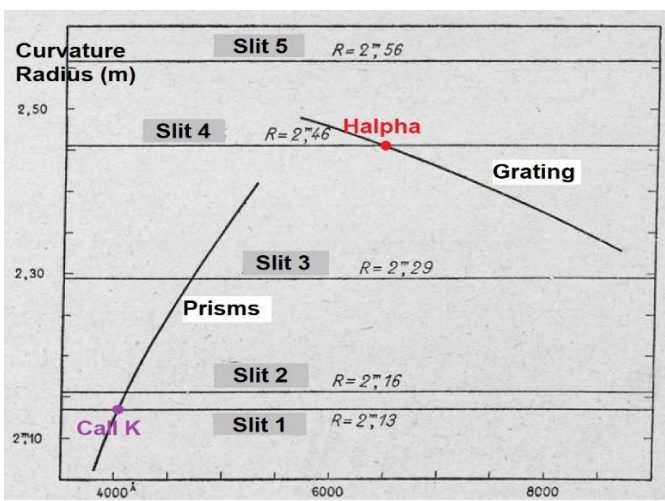

**Figure 7.** *The radius of curvature of spectral lines as a function of wavelength for the 7 metres research chamber n° IV. The radius does not vary much with the grating, but this is not the case with the prisms in the blue/violet part of the spectrum. Five 22 cm long slits were available (slits 1, 2, 3 for the blue, slit 4 for the green and red).*
*After d'Azambuja (1930).*

Without any reduction, the size of chamber n° IV was 7 metres; with image reduction, chamber n° IV was called n° IVbis and the length was 14 metres (figure 8). It was not possible to move very fast from chamber n° I (dedicated to CaII K) to chamber n° IV because two parts of chamber n° IV had to be dismounted when it was not used (the $F_4$ environment, which was suspended at the ceiling, figure 5, and the enclosure of chamber n° IVbis, figure 8). On the contrary, it was very easy to switch between chambers n° I, n° II (daily observations of CaII K and Hα) and n° III, which was a multi-purpose spectrograph of 3.0 m focal length, medium dispersion as shown by figure 4, working either with the 3 prisms or the grating, and dedicated to research.

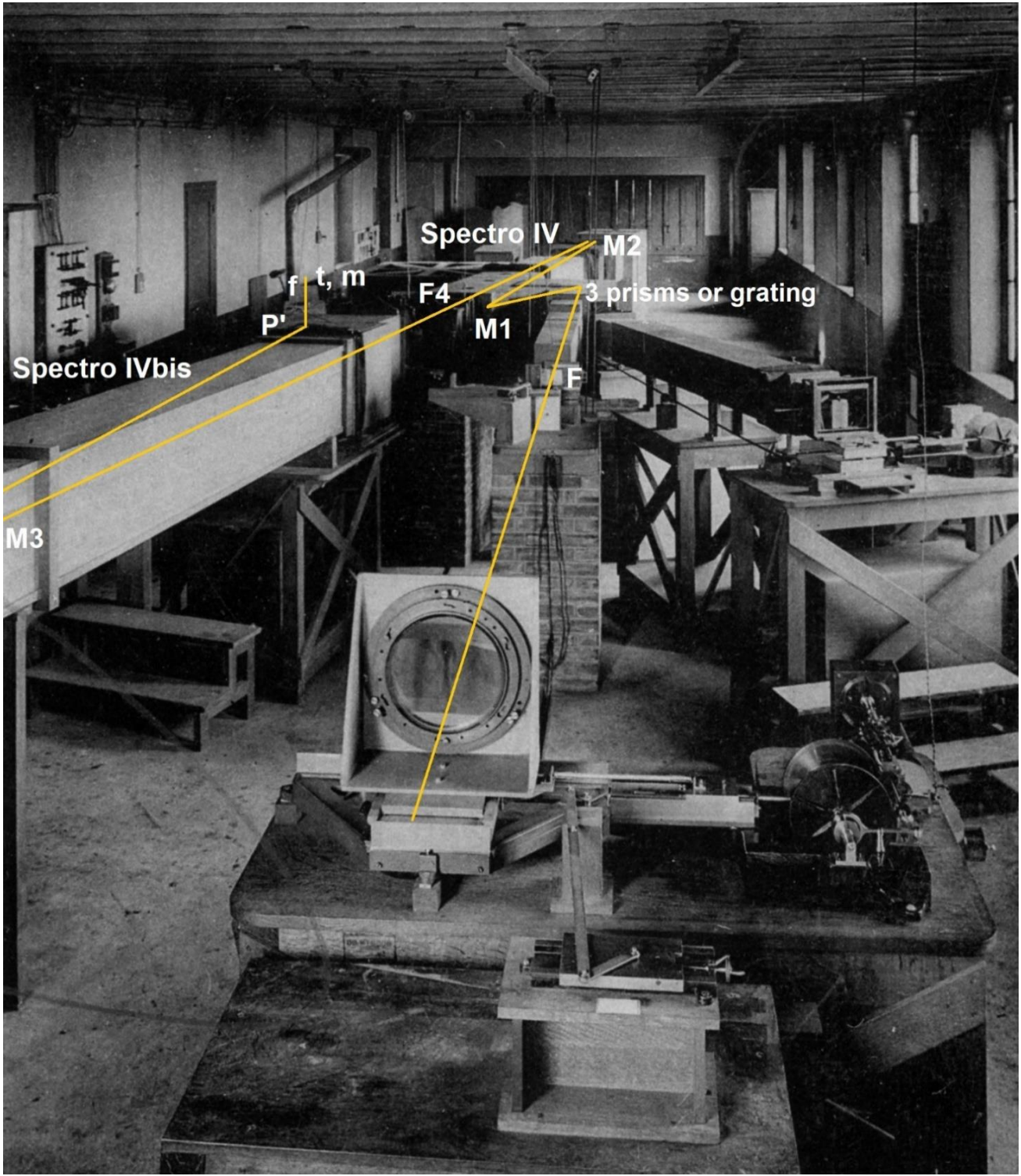

**Figure 8.** *The 2 x 7.0 meters chambers n° IV and n° IVbis. $M_1$ is a flat mirror; $M_2$ is the 7.0 m concave mirror of chamber n° IV, forming the spectrum on the slit $F_4$ (image size 200 mm). $M_3$ is the 7.0 m concave mirror of chamber n° IVbis; it forms an afocal system with one lens near prism P' to reduce the image size and focuses the output spectrum on the exit slit f. The distance between $M_2$ and $M_3$ is 14 m. The couple (t, m) is the motorized support of the photographic plate. Courtesy Paris Observatory.*

The chambers n° IV and n° IVbis were intensively used by d'Azambuja (1930) for his thesis work and observations of unusual lines. For example, figure 9 shows spectroheliograms in the infrared lines of CaII (8498 Å and 8542 Å) which were compared to CaII K3 and K1v, in order to validate the initial choice of Deslandres for chromospheric imagery. This figure also shows the first spectroheliogram got worl-wide in HeI 10830 Å (filaments as well as plages, usually bright, appear dark). These observations were not reproduced later. However, daily observations in the infrared lines of CaII and HeI started in 1974 at Kitt Peak National Observatory. The CaII 8542 Å is convenient for chromospheric magnetic fields (Zeeman effect). EUV images of HeII 304 Å (80000 K temperature) are produced by EIT/SOHO since 1996 and AIA/SDO since 2010.

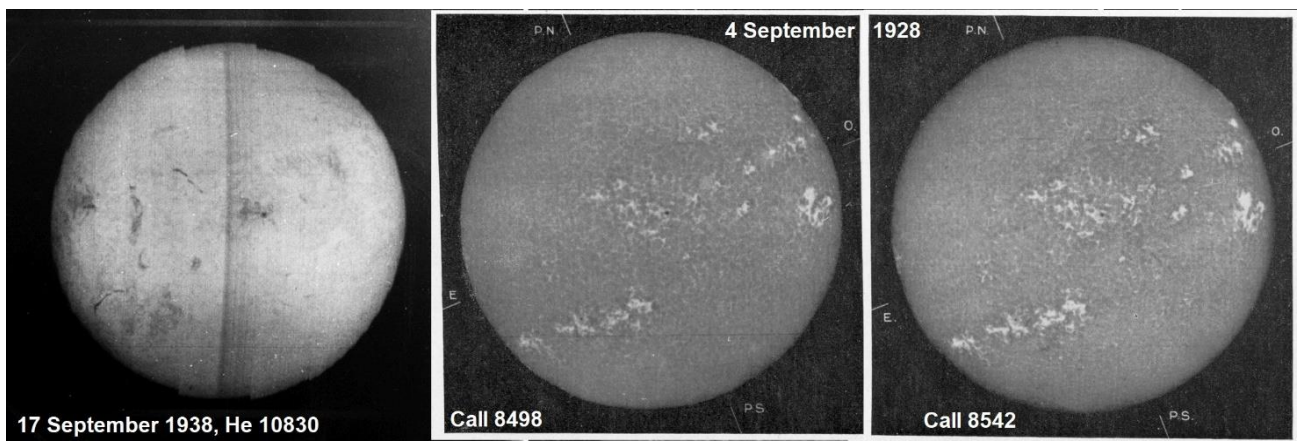

**Figure 9.** *Spectroheliograms in special lines made with chambers n° IV and n° IVbis. HeI 10830 Å (left, after d'Azambuja & d'Azambuja, 1938) and the two CaII infrared lines (right, after d'Azambuja, 1930).*

## 2 – THE SPECTROHELIOGRAPH OF THE MODERN ERA (1989 – TODAY)

The spectroheliographs n° III, n° IV and n° IVbis were dismantled in 1960 after the retirement of Lucien d'Azambuja (1954) and his wife Marguerite, born Roumens (1959). At the beginning of the eighties, after 70 years of continuous service, it was necessary to refurbish the optical and mechanical parts of the spectroheliographs n° I and n° II in order to prolongate systematic observations of CaII K and Hα. It took many years and the new version (figure 10) was ready on 1st January 1989.

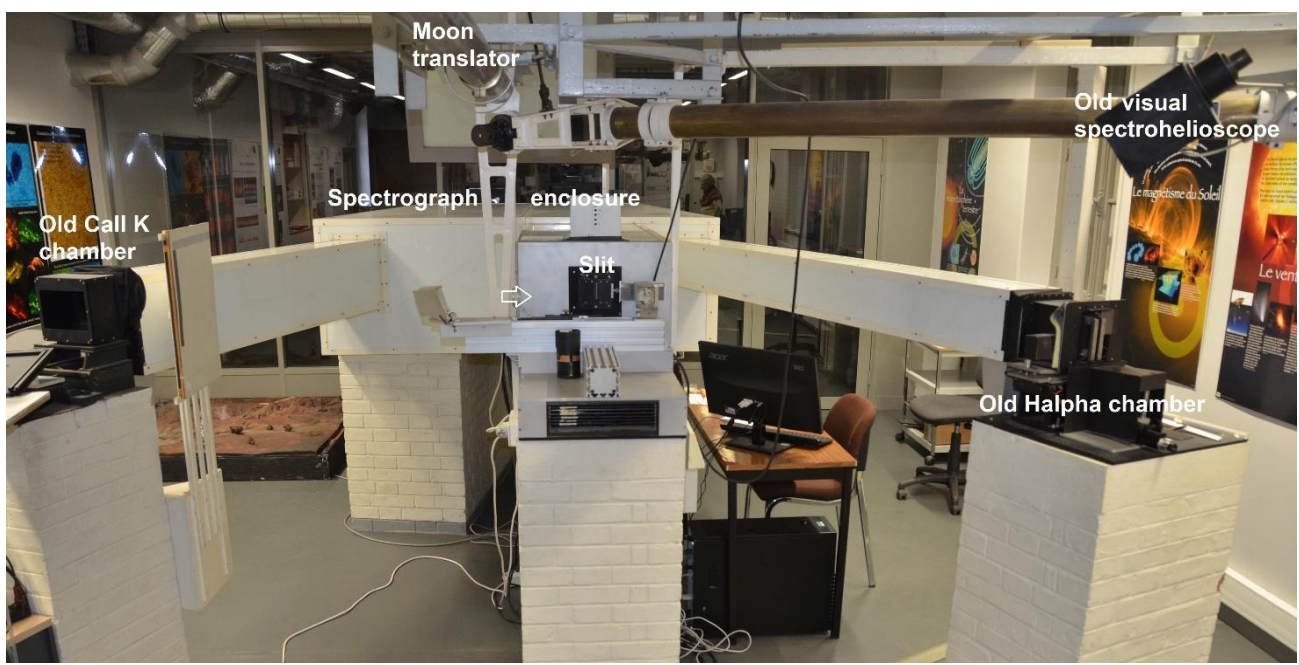

**Figure 10.** *The 1989 version of the spectroheliograph. Courtesy Paris Observatory.*

The 1989 version was studied by Gualtiero Olivieri (1928-2022). The entrance objective was replaced, together with the enclosures of the spectroheliographs. Optical elements of the spectrograph were also replaced, but the overall capabilities of the instrument were saved, for compatibility reasons with the glass plates collection. Chamber n° I (left, figure 10) was no longer used, but preserved, while chamber n° II (right, figure 10) was dedicated to systematic observations of CaII K and Hα. For that purpose, the old set of 3 prisms and plane grating were replaced by a 300 grooves/mm blazed grating (b = 17°27') working at order 3 for Hα and order 5 for CaII K. The grating was fixed and, for both lines, it worked with incidence and diffraction angles of respectively 7° and 27°. The interference orders were selected by coloured filters (figure 11). The focal length of the chamber (3.0 m) did not change, but glass plates were replaced by 13 x 18 cm² film plates. The dispersion was 0.30 mm/Å and 0.50 mm/Å, respectively for Hα and CaII K. With a 0.075 mm selecting slit in the spectrum, the bandpass was 0.25 Å and 0.15 Å, respectively for Hα and CaII K. Hβ was available in order 4 but not observed, because of the chromatism of the optics which was optimized for the red and the violet.

In order to observe together prominences at the limb and filaments on the disk, a disk attenuator (figure 12) was introduced to increase the exposure time without burning the disk. It is an artificial moon of neutral density 1.0 which moves upon the solar image (37.2 mm diameter) at the focus of the entrance objective. Two moons are available, one for winter and another one for summer. Standard observations consisted of CaII K3, CaII K1v, Hα, without any attenuator, and a long exposure CaII K3 with the attenuator to reveal prominences.

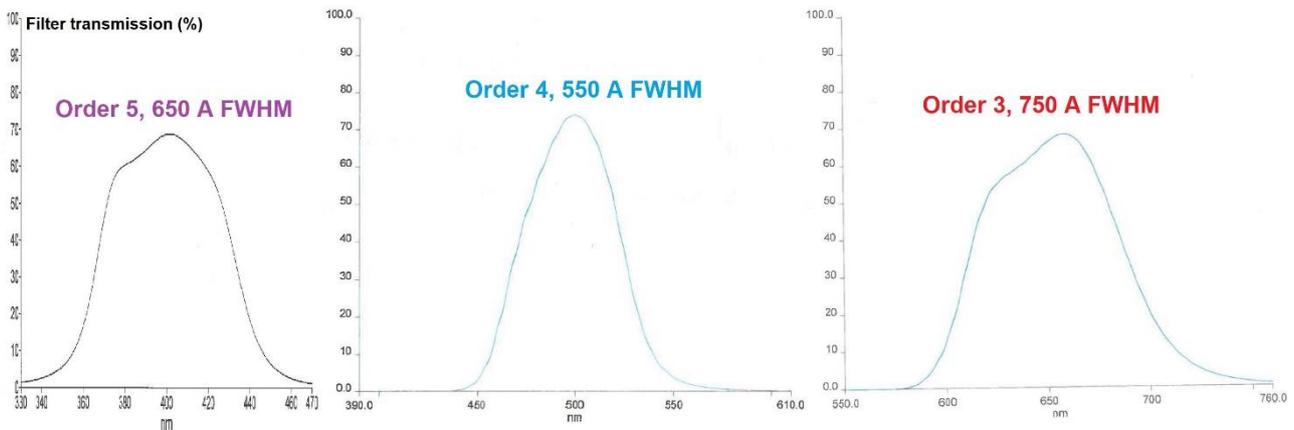

**Figure 11.** *Bandpass of selecting order filters (wavelength in abscissa, in nanometres; transmission in ordinates, in %). Courtesy Paris Observatory.*

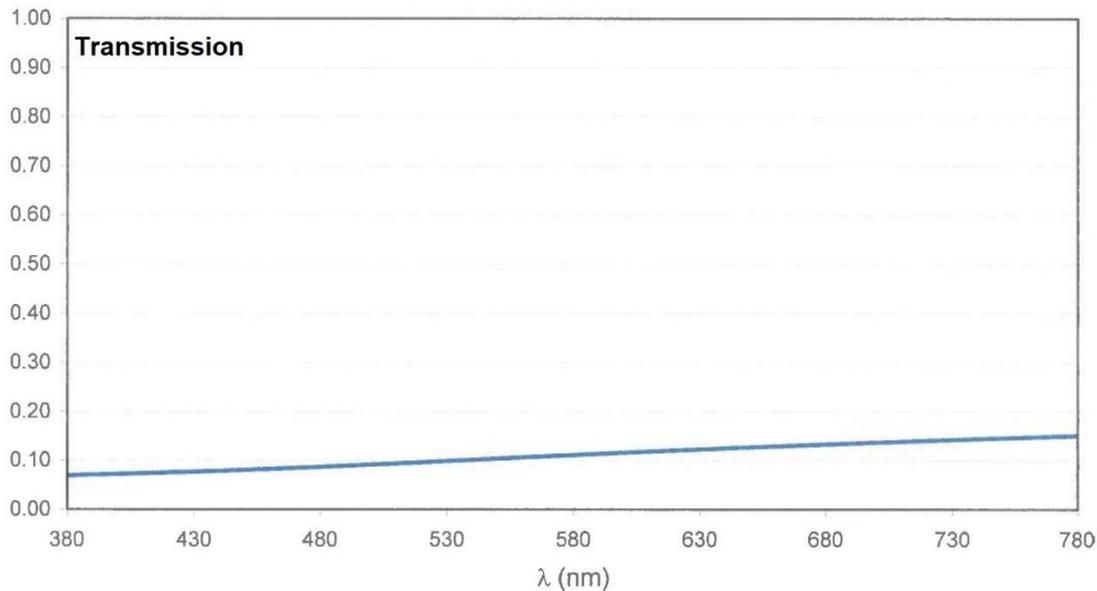

**Figure 12.** *Transmission of the disk attenuator (wavelength in abscissa, in nanometres; transmission in ordinates, in %). Courtesy Paris Observatory.*

The 1989 version was also ready for testing a Reticon linear array of 2048 pixels (13 µ square) located in another 1.0 m focal length chamber inside the spectrograph. With a linear array, the curvature of spectral lines (radius 1.55 m for 1.0 m focal length, wavelength independent at constant angles) must be suppressed; for that purpose, the entrance slit of the spectrograph was replaced by a curved slit of radius 1.965 m, in order to compensate the curvature of the spectrum. This change introduced a small geometric distorsion in the final image with a maximum shift of 4.5" at the poles (2.2" at 45° latitude, 1.1" at 30° latitude, null at the equator). The Reticon focus was never used for regular observations, because rectangular CCD appeared soon.

The next version is the one of 2002. It was also a major release, because the photographic chambers were abandoned (but preserved for historical reasons), and the selecting slit in the spectrum was suppressed. The 2002 version uses a specific 0.90 m chamber, located inside the spectrograph (figure 13), and a back illuminated CCD from Princeton Instruments (1340 x 100 pixels, 1 MHz, 20 µ square pixels, 200000 electrons full well, 14 bits dynamic range, 6-9 electrons of readout noise, water cooled at -40°C operating temperature, ST138 controler). As the CCD was too slow, a 1340 x 5 pixels window was recorded, so that the core of line profiles was described by 5 points only. The spectral sampling was 0.22 Å and 0.13 Å, respectively for Hα and CaII K. The quantum efficiency was exceptional (figure 14) both in orders 3 and 5.

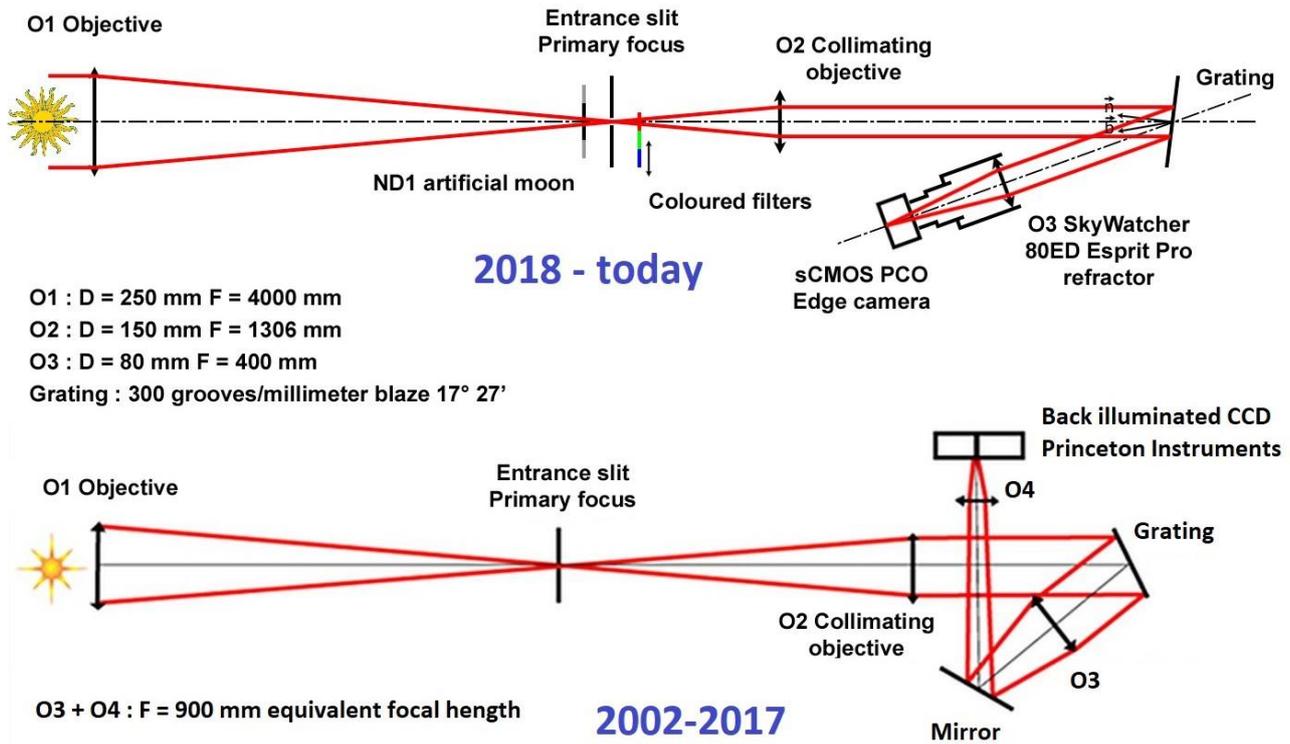

**Figure 13.** *The numerical versions of the spectroheliograph. Bottom: 2002 version with CCD; top: 2018 version with scientific CMOS. Courtesy Paris Observatory.*

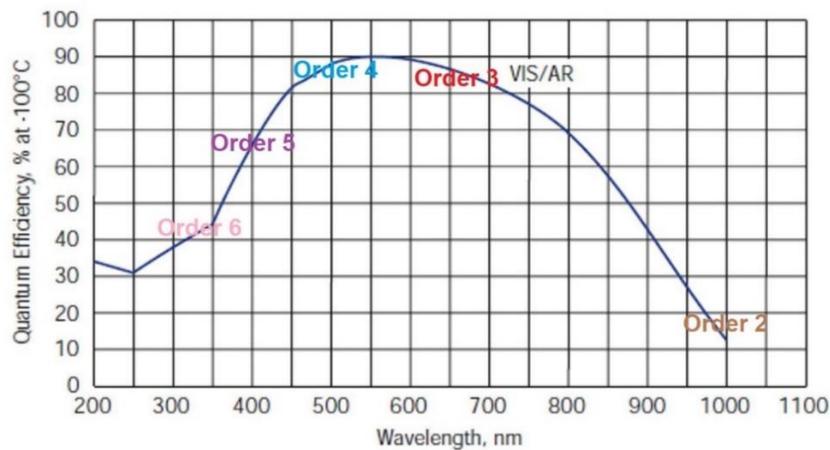

**Figure 14.** *The quantum efficiency of the 2002 back illuminated CCD detector with VIS/AR + Lumogen coating. CaII K and Hα are respectively observed in orders 5 and 3. Hβ is visible in order 4. Courtesy Paris Observatory.*

Figure 15 shows an example of daily observations made since 1989, on film plates until 2001 and with the CCD after. Hα, CaII K3, CaII K1v and a long exposure CaII K3 for prominences were done two or three times per day, according to weather conditions. The figure shows the famous day of 28 October 2003, a huge X17-class flare delopped in AR0486. The spatial sampling of the CCD was 1.7"/pixel after 2001.

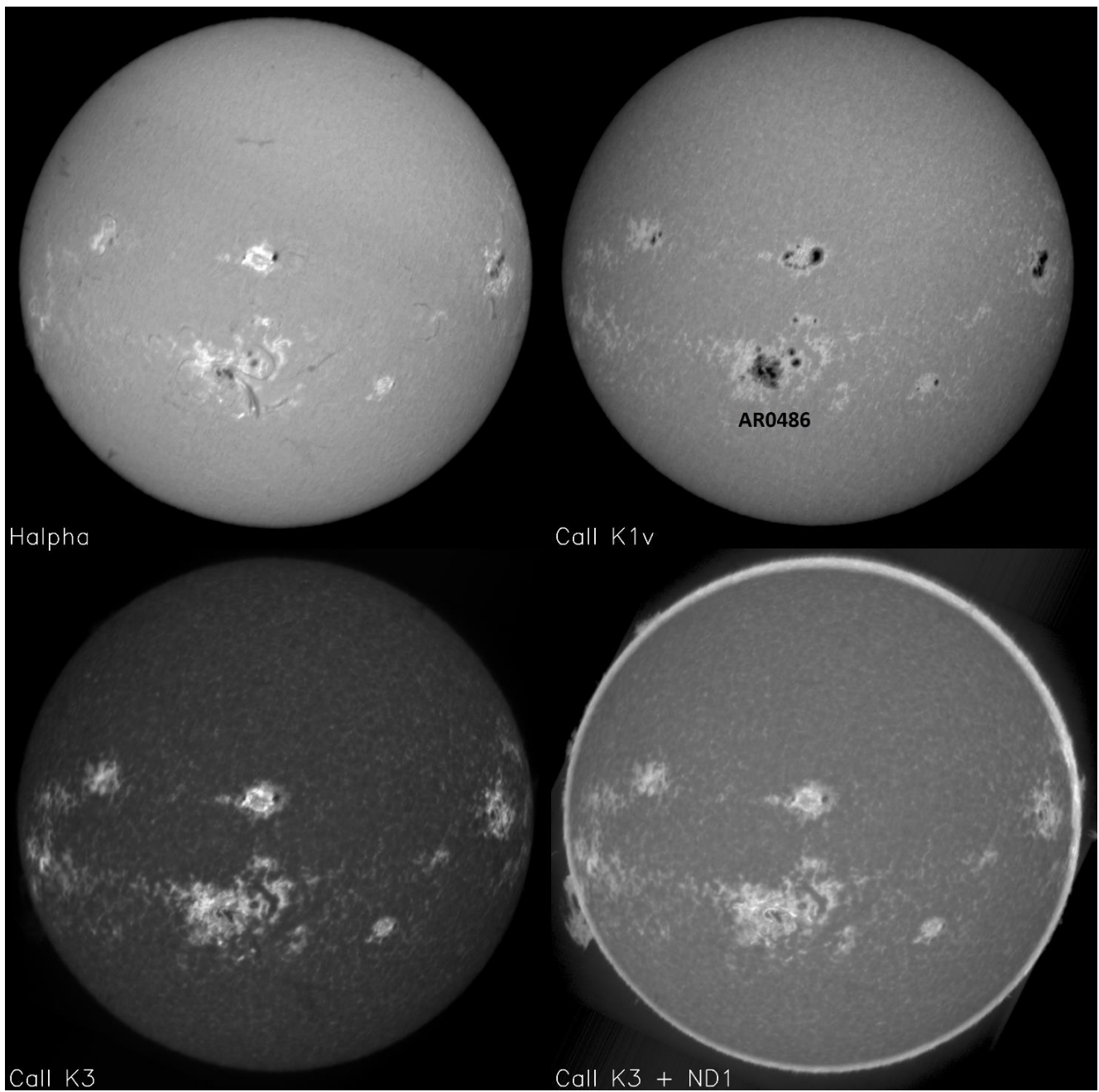

**Figure 15.** *Example of series done daily since 1989. Hα and CaII K3 for the chromosphere, CaII K1v for the photosphere, and a long exposure CaII K3 with a disk attenuator for prominences. 28 October 2003. First CCD version of the spectroheliograph. Courtesy Paris Observatory.*

The last version appeared in July 2017. It was also an important event, because for the first time, full line profiles were recorded for each solar pixel, providing (x, y, λ) FITS data-cubes. The classical images are slices of the cube. This version uses a specific 0.40 m chamber, located inside the spectrograph (figure 13), and a scientific CMOS array (PCO camera, Fairchild CIS2020 sensor, 2048 x 2048 pixels, 400 MHz, 6.5 µ square pixels, 30000 electrons full well, 16 bits dynamic range, 1.4 electrons of readout noise, water cooled at 5°C operating temperature). We record 40 points along the Hα line profile (6 Å wide range), and 100 points along the CaII H and CaII K line profiles (9 Å wide range) which are observed simultaneously (figure 16 and figure 17). The spectral sampling is 0.155 Å and 0.093 Å, respectively for Hα and CaII H/K. The quantum efficiency is excellent (figure 18) in orders 3, 4 (Hydrogen lines) and moderate in order 5 (CaII). The wavelength sampling of spectral lines is reported in figure 19. The spatial sampling is now 1.1"/pixel.

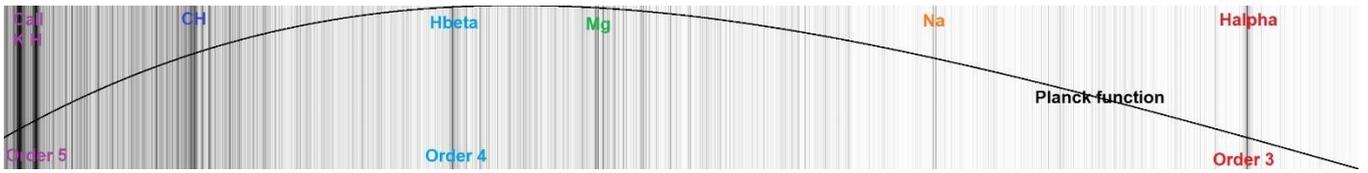

**Figure 16.** *Full spectrum from 3900 Å to 6800 Å and interference orders. Courtesy Paris Observatory.*

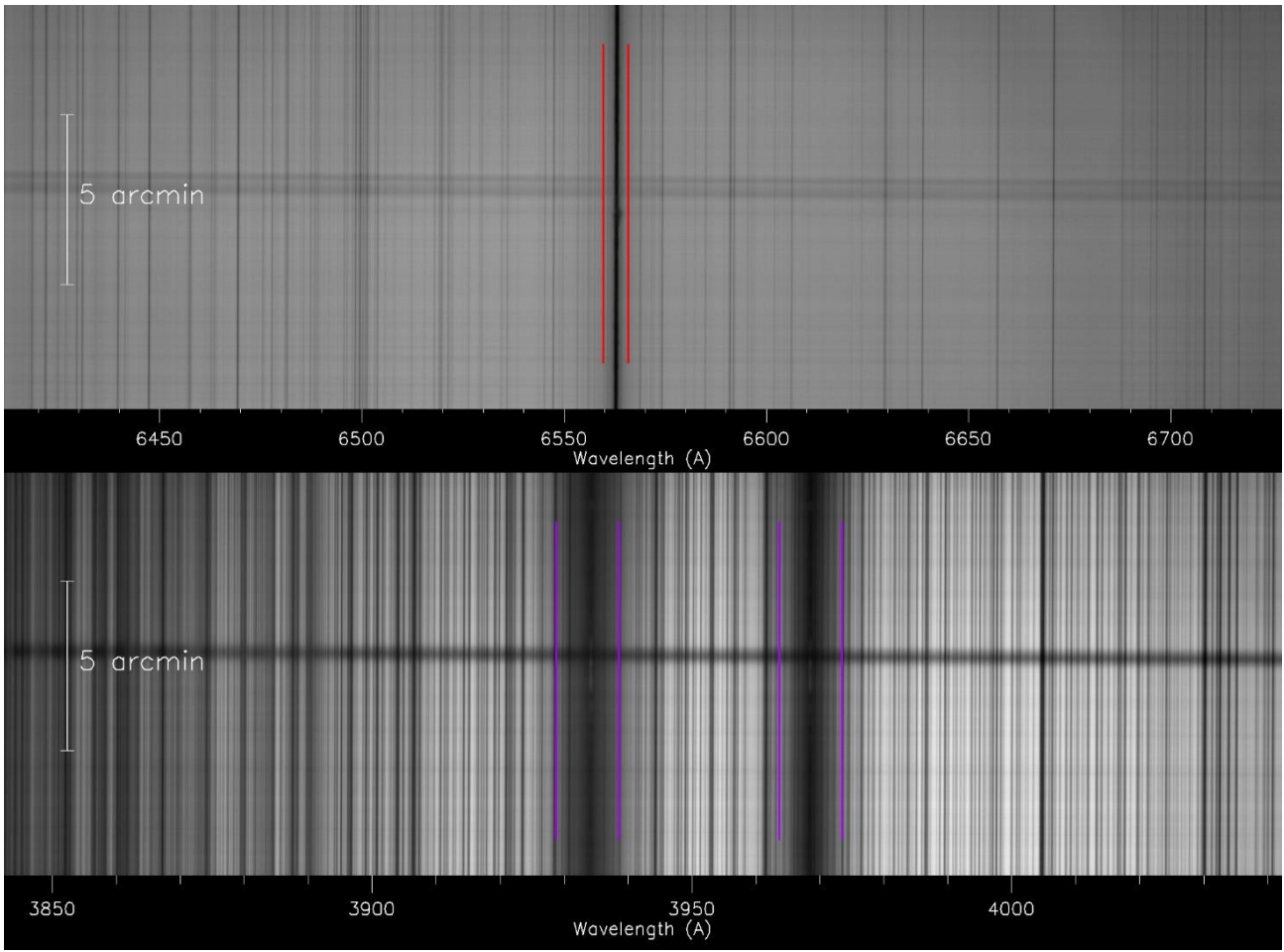

**Figure 17.** *Windows which are selected in the observed spectrum to record the line profiles of Hα (top) and CaII K and H (bottom). CaII K and H are recorded simultaneously. Courtesy Paris Observatory.*

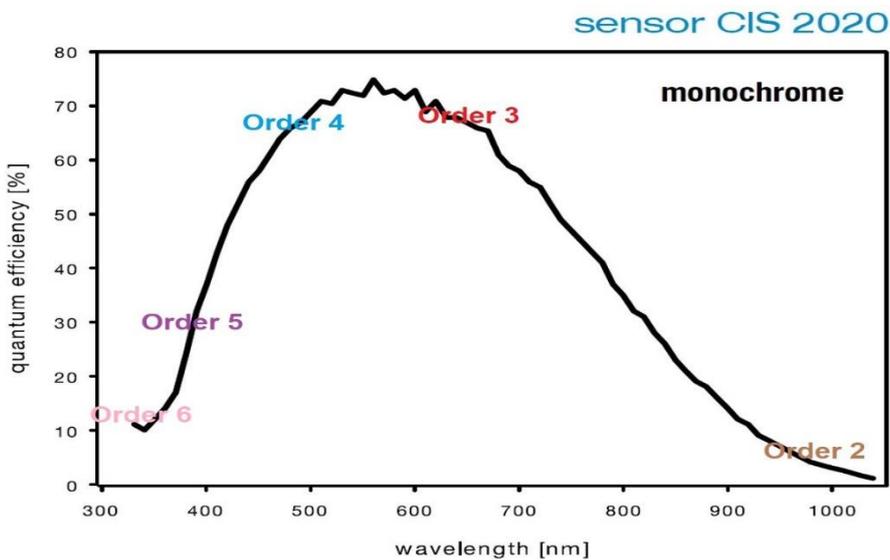

**Figure 18.** *Quantum efficiency (in %) of the CIS2020 sCMOS detector as a function of wavelength. Courtesy Paris Observatory and PCO.*

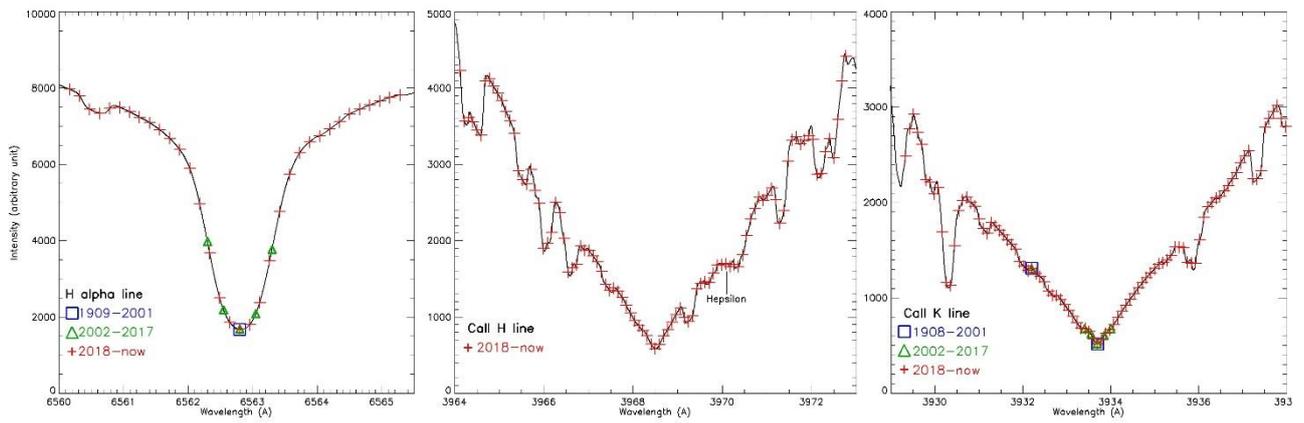

**Figure 19.** *Evolution of the spectral sampling of the line profiles of Hα, CaII K and H. 1909-2001: one point (squares). 2002-2017: 5 points (triangles). 2018-now: 40 to 100 points (red crosses). The CaII H collection is new: it started in July 2017. CaII K and H are observed simultaneously. Courtesy Paris Observatory.*

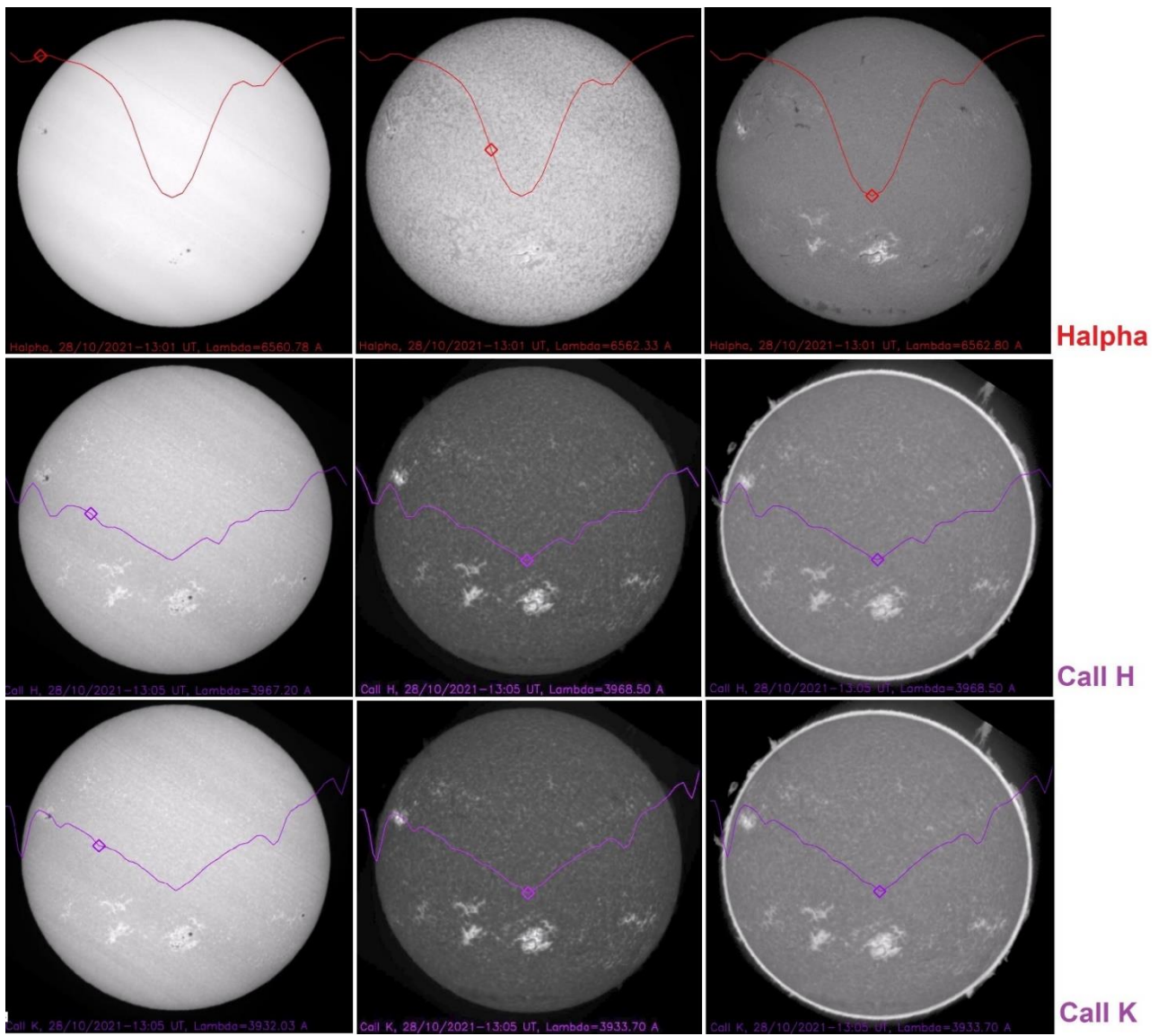

**Figure 20.** *Example of daily observations since July 2017. Top: Hα line (40 points in the profile), the selected images are the blue continuum, the blue wing and the line core. Centre: CaII H line (100 points in the profile), the selected images are the blue wing H1v and line core H3 (short and long exposure). Bottom: the same for CaII K line with K1v and K3. 28 October 2021. Courtesy Paris Observatory.*

Figure 20 shows an example of observations performed since July 2017 (here 28 October 2021). Up to three series of four observations are performed daily, with Hα, Hα long exposure for prominences, CaII H/K simultaneously, and CaII H/K long exposure with the disk attenuator. Each observation is a data-cube (x, y, λ), twelve cubes/day at maximum, from which classical images (slices) are derived at fixed wavelength.

Table 3 summarizes the capabilities of the modern versions of the spectroheliograph since 1989. We plan to replace the 6.5 µm sCMOS camera by a more recent model with smaller pixels (4.6 µm) without any change of optics. This will improve both the spectral sampling and spatial sampling (from 1.1" to 0.78", in comparison to the usual Meudon seeing of 2"). The optical spectral resolution, with an entrance slit of 0.03 mm (1.55") is 0.25 Å and 0.15 Å respectively for Hα and CaII H/K. With a 4.6 µm pixel, the sampling would be well optimized (accordingly to the Shannon theorem), with 0.11 Å and 0.07 Å, respectively for Hα and CaII H/K.

**Table 3**. *Features of the modern versions of Meudon spectroheliograph. BW = bandwidth of monochromatic images. The focal length of the collimator is 1.30 m. The right column indicates the number of simultaneous wavelengths recorded along the line profiles (for electronic detectors only).*

| Date | Chamber (m) | Hα BW (Å) | CaII K BW (Å) | Hα dispersion (Å/mm) | CaII K dispersion (Å/mm) | Wavelengths (pixels) |
|---|---|---|---|---|---|---|
| 1989 | 3.0 | 0.25 | 0.15 | 3.2 | 1.9 | 1 |
|  | slit 75 µm, FILM |  |  |  |  |  |
| 2002 | 0.90 | 0.22 | 0.13 | 12.5 | 7.5 | 5 |
|  | CCD 20 µm |  |  |  |  |  |
| 2018 | 0.40 | 0.155 | 0.093 | 28.3 | 14.3 | > 100 |
|  | sCMOS 6.5 µm |  |  |  |  |  |
| 2024 ? | 0.40 | 0.114 | 0.068 | 28.3 | 14.3 | > 150 |
|  | sCMOS 4.6 µm |  |  |  |  |  |

## 3 – A PRO/AM COLLABORATIVE PROJECT: OBSERVATIONS WITH THE SOLAR EXPLORER (SOLEX), A MINI- SPECTROHELIOGRAPH (2023)

In order to improve the temporal coverage of solar observations freely available to the international community (a maximum of 250 days/year are observed at Meudon), we have initiated a collaboration with amateur astronomers. A mini-spectroheliograph was proposed to them by Christian Buil in 2022, the SOLar Explorer (SOLEX, http://www.astrosurf.com/solex/sol-ex-presentation-en.html) together with a sophisticated processing software (INTI) written by Valérie Desnoux (http://valerie.desnoux.free.fr/inti/), with capabilities close to the Meudon instrument. Amateur data are now delivered by the BASS2000 solar archive (https://bass2000.obspm.fr/home.php) since February 2023, together with professional data. Many SOLEX are disseminated world-wide. Figure 21 shows an example of SOLEX at the focus of a Vixen apochromatic refractor (ED 81 mm / 625 mm) with focal reducer 0.67, designed for full Sun observations with ZWO ASI178.

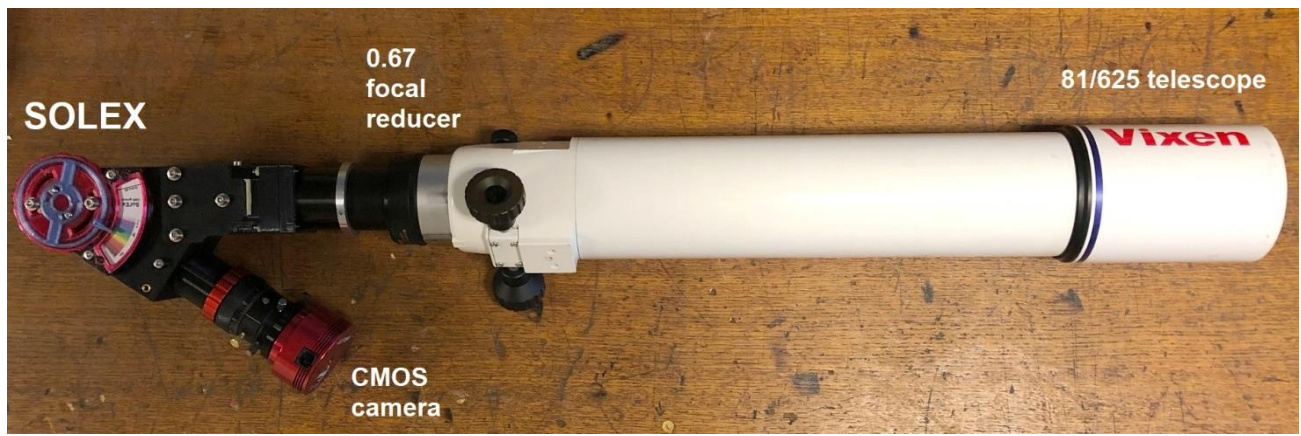

**Figure 21**. *Example of SOLEX instrument. Courtesy Paris Observatory.*

The SOLEX instrument is a compact spectrograph (figure 22) with a 4.5 mm slit (0.01 mm wide) that is compatible with equivalent focal lengths in the range 380-480 mm for full disk observations (the median focus of 430 mm provides an image of 4.0 mm on the slit). The spectrograph is produced by 3D-printers; the focal lengths of the collimator and chamber lenses are, respectively, 80 mm and 125 mm, so that the solar diameter on the detector is about 6 mm. The dispersive device is a 2400 grooves/mm grating at order 1. All optical elements are provided by Shelyak Instruments, and the standard camera is the CMOS ZWO ASI178 monochrome model (3096 x 2080 pixels) with 2.4 µ square pixels (14 bits). The quantum efficiency (figure 23) is about 50% for Hα and CaII K. However, the spectrograph can select any line in the range 3500–7500 Å. The dispersion is 0.038 mm/Å and 0.031 mm/Å, respectively for Hα and CaII K. For a focal length of 430 mm, the width of the slit corresponds to about 4.8".

The spectral resolution, sampling and miscellaneous informations are given by table 4 for a typical SOLEX designed for full Sun observations (430 mm focal length and ZWO ASI178 camera) and Meudon spectroheliograph, allowing to compare some of their main characteristics. Figure 24 shows two Hα spectra got with both instruments with the pixel sampling of 0.063 Å and 0.155 Å, respectively for SOLEX and Meudon. Two Hα images are displayed in figure 25 and demonstrate the high quality of the SOLEX instrument.

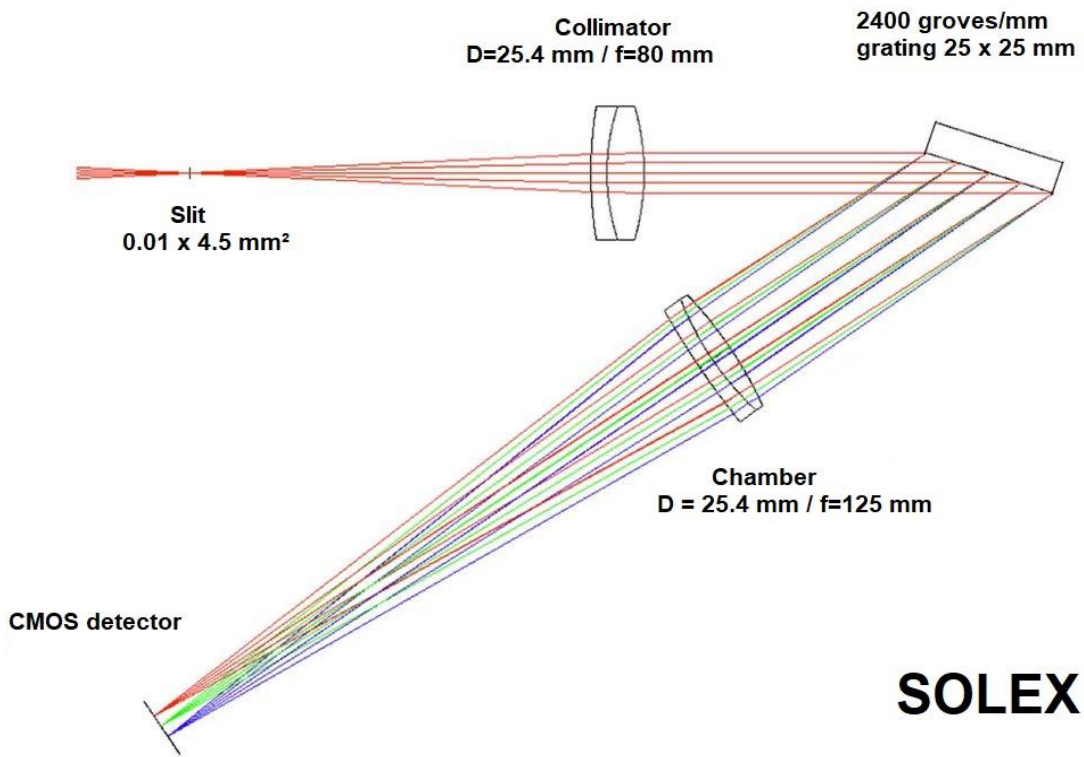

**Figure 22.** *The SOLEX optical design. Courtesy Christian Buil.*

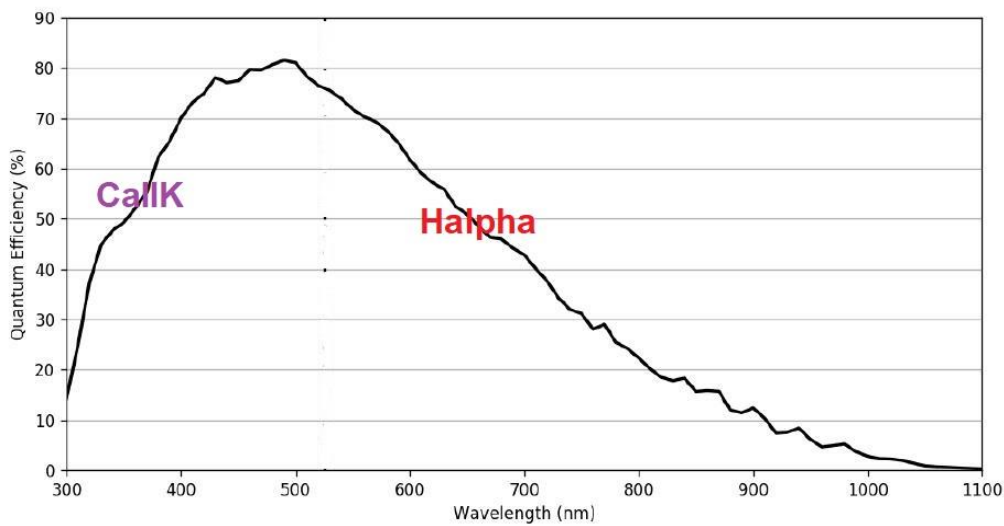

**Figure 23.** *The IMX178 quantum efficiency (%). Courtesy ZWO.*

**Table 4.** *Main characteristics of SOLEX and the 2017 version of Meudon spectroheliograph (SHG).*

| | | SOLEX | Meudon SHG | Remarks |
|---|---|---|---|---|
| SPECTRO | Spectral resolution | 0.16 Å | 0.25 Å | Hα |
| | Spectral resolution | 0.36 Å | 0.15 Å | CaII H/K |
| | Spectral sampling | 0.063 Å | 0.115 Å | Hα |
| | Spectral sampling | 0.077 Å | 0.093 Å | CaII H/K |
| | Slit width | 4.8" (10 μ) | 1.55" (30 μ) | |
| | Slit height | 4.5 mm | 45 mm | |
| | Spatial sampling | 1.15" | 1.10" | |
| | Radius of line curvature | 6-13 cm | ∞ | SHG: curved entrance slit |
| | Grating | 2400 gr/mm | 300 gr/mm | SOLEX: rotating grating |
| | Interference order | 1 | 3 and 5 | SHG: fixed grating |
| | Order selection | | filters | |
| OPTICS | Objective (dia./focal length) | 0.06/0.4 m | 0.25/4.0 m | SHG: 0.17 m diaphragm |
| | Collimator (dia./focal length) | 0.025/0.08 m | 0.13/1.3 m | |
| | Chamber (dia./focal length) | 0.025/0.125 m | 0.08/0.4 m | Old SHG: 3.0 m, 0.9 m |
| DETECTOR | Type | CMOS | Cooled sCMOS | Old SHG: Glass, film, CCD |
| | Pixel size | 2.4 μ | 6.5 μ | |
| | Full Well (electrons) | 15000 | 30000 | |
| | Readout noise (electrons) | 2.2 | 1.4 | |
| | Dynamic range | 6800 | 21400 | |
| MOUNT | type | equatorial | coelostat | SHG : fixed spectrograph |
| | scanning device | mount | objective | |

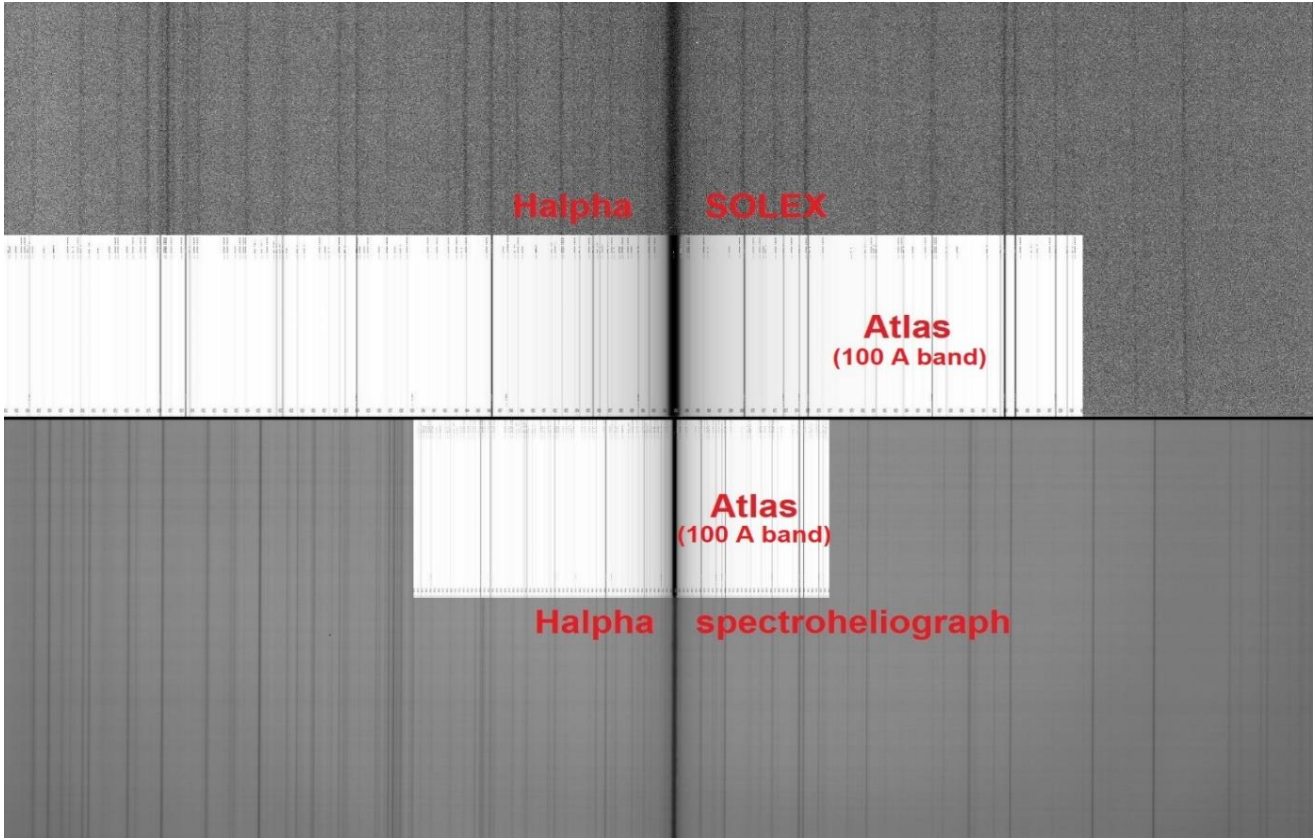

**Figure 24.** *Spectra of the Hα line got with SOLEX (top, 0.063 Å sampling) and Meudon spectroheliograph (bottom, 0.155 Å sampling). The noise on the top image is due to the presence of dense clouds during the observational test reducing considerably the photometry. Courtesy Paris Observatory.*

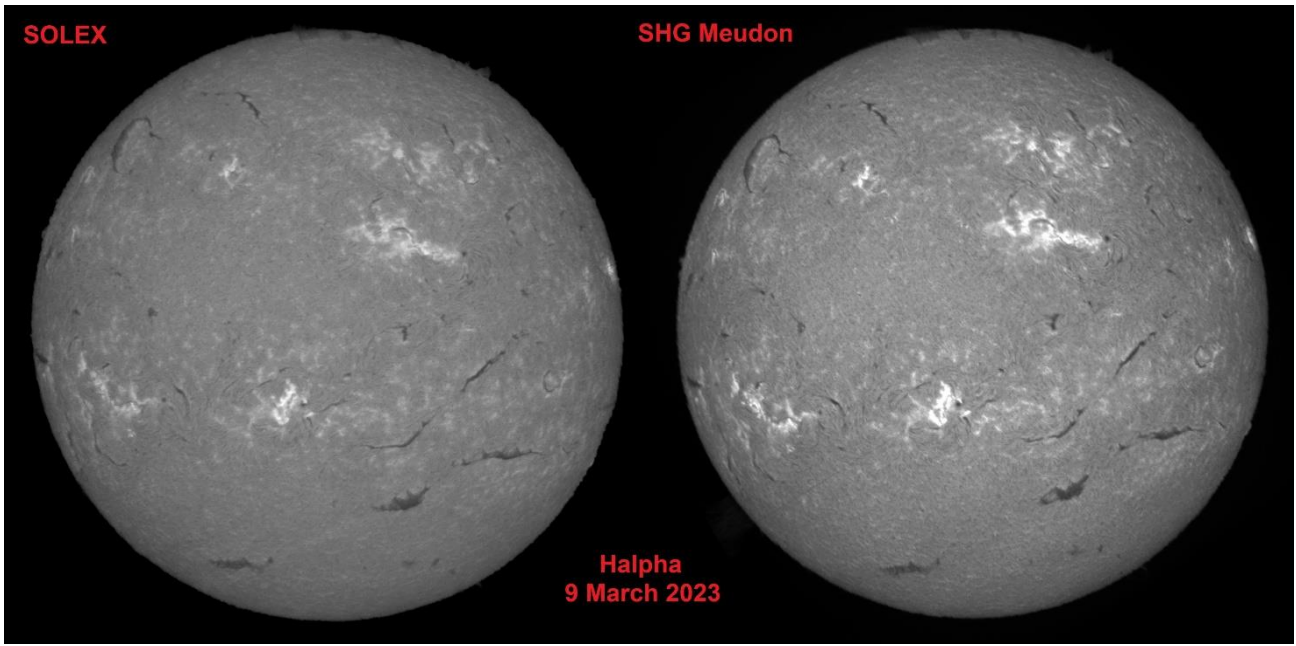

**Figure 25.** *Spectroheliograms in the centre of the Hα line got with SOLEX (left) and Meudon spectroheliograph (right) on 9 March 2023. For SOLEX: courtesy Eric Bernard, Saint Germier Observatory (Gers, France). For Meudon: courtesy Paris Observatory.*

The SOLEX instrument is able to observe a large variety of lines (either photospheric or chromospheric) in the range 3500-7500 Å, just by rotating the grating at order 1, as shown by figure 26. This capability is of great interest to explore other lines than the usual Hα or CaII H/K, and can be used for more sophisticated observations, such as magnetometry of the photosphere in some Zeman sensitive lines. The dispersion lies in the range 0.03-0.05 mm/Å and increases towards the red. The curvature of spectral lines also increases towards the red (the radius decreases with wavelength and lies in the range 4-13 cm). On the contrary, the grating of Meudon spectroheliograph is fixed, so that only a few lines appearing in orders 3 to 5 are theoretically observable. The line curvature does not depend of orders and is compensated by the curvature (2 m radius) of the entrance slit. In practice, the Meudon instrument is limited to orders 3 (Hα) and 5 (CaII H/K), because the objectives are optimized only for these orders and present chromatism outside. For instance, Hβ at order 4 is possible, but one has to correct the focal length variation of the imaging objective.

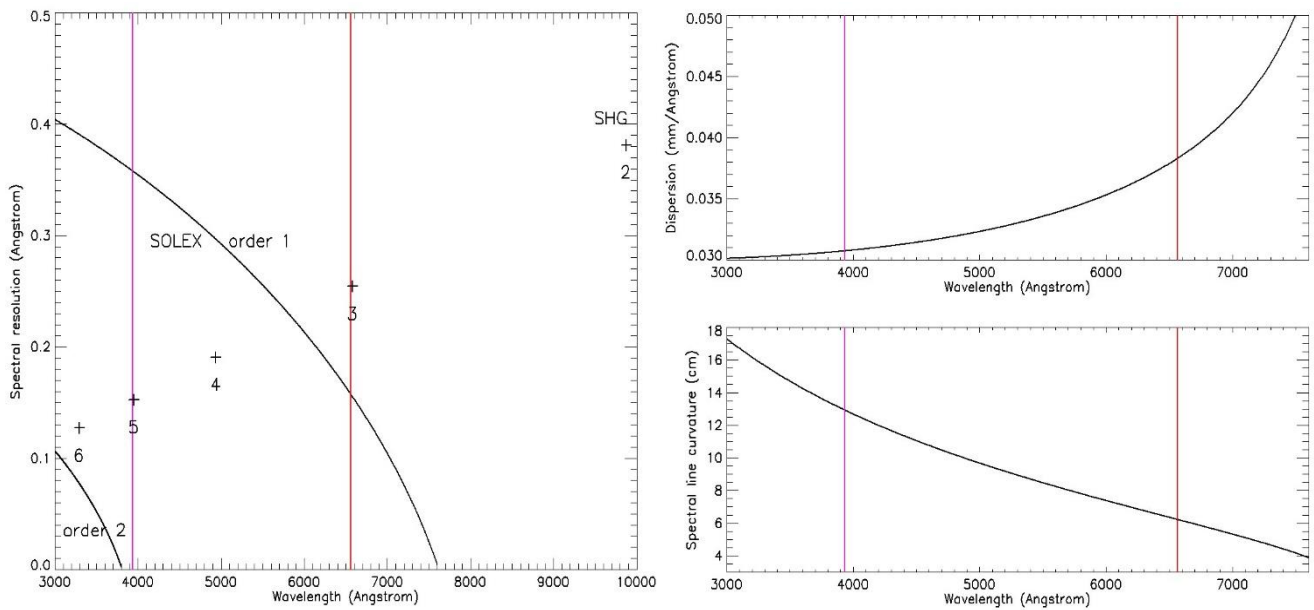

**Figure 26.** *Left: the spectral domain that can be explored with SOLEX or with Meudon spectroheliograph (which is restricted to discrete wavelengths in orders 3 to 5). Right: dispersion (mm/Å) and line curvature radius (cm) for SOLEX. Hα and CaII H/K lines are indicated by the coloured bars. Courtesy Paris Observatory.*

## 4 – CONCLUSION

We described in this paper the evolution of the optical design of Meudon Spectroheliograph since 1908. Most optical elements were revisited in the eighties. The instrument was optimized for producing monochromatic images of the photosphere and the chromosphere in the Hα and CaII H/K lines, and for that reason, it is not a universal spectrograph. The photographic version used glass and film plates until 2001. Since 2002, they were replaced by a CCD; in 2017, a fast sCMOS detector was integrated to the spectrograph and allows, for the first time, to record the full line profiles for each pixel of the Sun, under the form of (x, y, λ) data-cubes. The image collection covers 115 years of solar activity. A collaboration with amateur astronomers is starting (2023) with many compact (but high quality) spectroheliographs (SOLEX project) disseminated in several places. We hope for observations 365 days/year and a fruitful cooperation between amateurs and professionals in solar physics.

## 5 - REFERENCES

## 6 - ON-LINE MATERIAL (VIDEO CLIPS, FIGURES)

Figures:
https://drive.google.com/drive/folders/1XmBDSVfQAsJXxikIEz_0dT8_Up9fi7Cw?usp=sharing

Movie 1: scans of the solar surface with Meudon Spectroheliograph
https://drive.google.com/file/d/1wZOnkvBJYfa3nWb4dliL6fK7vIRykF1Q/view?usp=share_link

Movie 2: wavelength scans of Hα, CaII H and K spectral lines with Meudon Spectroheliograph
https://drive.google.com/file/d/1l3Rzm-HhX4K5Hgxes__lrTw0guK0NOLw/view?usp=share_link

## 7 - ACKNOWLEDGEMENTS

The author thank I. Bualé and F. Cornu for spectroheliograms and archives of the Meudon solar collection, Ch. Buil for the SOLEX documentation, E. Bernard for the SOLEX image of figure 25 and D. Crussaire for spectroscopic tests.
## 8 - THE AUTHOR

Dr Jean-Marie Malherbe, born in 1956, is astronomer at Paris-Meudon observatory. He got the degrees of "*Docteur en astrophysique*" in 1983 and "*Docteur ès Sciences*" in 1987. He used the spectrographs of the Meudon Solar Tower, the Pic du Midi Turret Dome, the German Vacuum Tower Telescope, THEMIS (Tenerife) and developed polarimeters. He also used the space-born instruments HINODE (JAXA), SDO and IRIS (NASA). He is responsible of the Meudon spectroheliograph since 1996.